\newread\testifexists
\def\GetIfExists #1 {\immediate\openin\testifexists=#1
    \ifeof\testifexists\immediate\closein\testifexists\else
    \immediate\closein\testifexists\input #1\fi}
\def\epsffile#1{Figure: #1}     

\immediate\openin\testifexists=e
    \ifeof\testifexists\immediate\closein\testifexists\else
    \immediate\closein\testifexists\input e\fipsf

\magnification= \magstephalf \tolerance=1600
\parskip=5pt
\baselineskip= 4.5 true mm \mathsurround=1pt
 \hsize=5.3 in
 \vsize=7.6 in
\font\smallrm=cmr8

\font\medrm=cmr9

\font\bigbf=cmbx12
    \def\Bbb#1{\setbox0=\hbox{$\tt #1$}  \copy0\kern-\wd0\kern .1em\copy0}
    \immediate\openin\testifexists=a
    \ifeof\testifexists\immediate\closein\testifexists\else
    \immediate\closein\testifexists\input a\fimssym.def 
\def\secbreak{\vskip12pt plus 1in \penalty-200\vskip 0pt plus -.8in}
   \def\newsect#1{\secbreak\noindent{\bf #1}\medskip}
\def\hugeskip{\vskip12mm plus 3mm}
\def\Narrower{\par\narrower\noindent}   
\def\Endnarrower{\par\leftskip=0pt \rightskip=0pt}
\def\br{\hfil\break}    \def\ra{\rightarrow}
\def\a{\alpha}      \def\b{\beta}   \def\g{\gamma}  
\def\d{\delta}        \def\e{\varepsilon}
          \def\l{\lambda}     
                     
         \def\j{\psi}    
\def\r{\varrho}     \def\s{\sigma}

\def\w{\omega}        

  \def\OO{{\cal O}}

\def\cl{\centerline}    
\def\ni{\noindent}      \def\pa{\partial}   \def\dd{{\rm d}}
            \def\ket{\rangle}

\def\fn#1{\ifcase\noteno\def\fnchr{*}\or\def\fnchr{\dagger}\or\def
    \fnchr{\ddagger}\or\def\fnchr{\medrm\S}\or\def\fnchr{\|}\or\def
    \fnchr{\medrm\P}\fi\footnote{$^{\fnchr}$}
    {\scrunch#1\toe}\ifnum\noteno>4\global\advance\noteno by-6\fi
    \global\advance\noteno by 1}
    \def\scrunch{\baselineskip=11 pt \medrm}
    \def\toe{\vphantom{$p_\big($}}
    \newcount\noteno

\def\ffract#1#2{{\textstyle{#1\over#2}}}
\def\fract#1#2{\raise .35 em\hbox{$\scriptstyle#1$}\kern-.25em/
    \kern-.2em\lower .22 em \hbox{$\scriptstyle#2$}}

\def\half{\ffract12} \def\quart{\fract14}

\def\part#1#2{{\partial#1\over\partial#2}}
 \def\ref#1{${\vphantom{)}}^#1$}
\def\ex#1{e^{\textstyle#1}}

\def\bbf#1{\setbox0=\hbox{$#1$} \kern-.025em\copy0\kern-\wd0
    \kern.05em\copy0\kern-\wd0 \kern-.025em\raise.0433em\box0}

\def\ref#1{${\,}^{\hbox{\smallrm #1}}$}

\def\Gbar{\raise.13em\hbox{--}\kern-.35em G}
\def\lap{\setbox0=\hbox{$<$}\,\raise .25em\copy0\kern-\wd0\lower.25em\hbox{$\sim$}\,}
\def\glt{\setbox0=\hbox{$>$}\,\raise .25em\copy0\kern-\wd0\lower.25em\hbox{$<$}\,}
\def\gap{\setbox0=\hbox{$>$}\,\raise .25em\copy0\kern-\wd0\lower.25em\hbox{$\sim$}\,}


{\ }\vglue 1truecm
 \rightline{SPIN-2000/07}\rightline{hep-th/0003005} \hugeskip
\cl{\bigbf Determinism and Dissipation in}\smallskip \cl{\bigbf
Quantum Gravity
}\hugeskip \cl{Gerard
't~Hooft }
\bigskip
\cl{Institute for Theoretical Physics}
\cl{University of Utrecht, Princetonplein 5}
\cl{3584 CC Utrecht, the Netherlands}
\smallskip
\cl{and}
\smallskip
\cl{Spinoza Institute}
\cl{Postbox 80.195}
\cl{3508 TD Utrecht, the Netherlands}
\smallskip\cl{e-mail: \tt g.thooft@phys.uu.nl}
\cl{internet: \tt http://www.phys.uu.nl/\~{}thooft/ }
\hugeskip
\ni{\bf Abstract}\Narrower
 Without invalidating quantum mechanics as a principle underlying the
 dynamics of a fundamental theory, it is possible to ask for even more
 basic dynamical laws that may yield quantum mechanics as the machinery
 needed for its statistical analysis. In conventional systems such as
 the Standard Model for quarks and leptons, this would lead to hidden
 variable theories, which are known to be plagued by problems such as
 non-locality. But Planck scale physics is so different from field
 theories in some flat background space-time that here the converse may
 be the case: we speculate that causality and locality can only be restored
 by postulating a deterministic underlying theory. A price to be paid may
 be that the underlying theory exhibits dissipation of information.
\Endnarrower
\hugeskip
\newsect{1. Introduction}
In the opening lecture of this School\ref1 it was explained that
the physical degrees of freedom of a black hole are distributed on
its horizon in such a way that there appears to be one Boolean
degree of freedom per unit of surface area of size $$A_0=4\,{\rm
ln}\,2\,L^2_{\rm Planck}\,.\eqno(1.1)$$ These are the degrees of
freedom that appear in the equations of motion, c.q. the
Schr\"odinger equation for a black hole. In turn, according to the
principles of General Relativity, these same equations of motion
should apply to what happens in the (nearly) flat space-time
experienced by an in-going observer. This led us to the
formulation of "the Holographic Principle", according to which the
physical degrees of freedom that describe all physical events in
some definite region of space and time, can be mapped onto a
two-dimensional surface in such a way that there is exactly one
Boolean degree of freedom per unit of surface of size $A_0$.

One striking aspect of this observation is that the same region of
space-time could be used to describe a different black hole with its
horizon some place else. Apparently, one may choose any member of an
infinite set of possible surfaces to project the physical degrees of
freedom onto it. Also, one should be able to map directly from one surface onto
another.

It now appears to be a satisfactory feature of string
theories\ref2 and certain semi-perturbative extensions of them,
that they manage to reproduce this holographic principle\ref3.
This principle implies that any complete theory combining quantum
mechanics with gravity should exhibit an upper limit to the total
number of independent quantum states that  increases exponentially
with the surface area of a system, rather than its volume.

All this raises a number of important questions. First, what does
{\it locality\/} mean for such theories? And how can notions such
as causality, unitarity, and local Lorentz invariance make sense
if there is no trace of `locality' left? In this lecture, a theory
is developed that will {\it not\/} postulate the quantum states as
being its central starting point, but rather classical,
deterministic degrees of freedom\ref4. Quantum states, being mere
mathematical devices enabling physicists to make statistical
predictions, may turn out to be derived concepts, with a not
strictly locally formulated definition. Once it is realized that
quantum states may be non-local, derived concepts, it is natural
to consider making one more step.

In the past, many versions of hidden variable theories were
dismissed by a majority of researchers for two reasons: one reason
was that these theories did not seem to work properly in the sense
that counter examples could be constructed using eigenstates of
certain symmetries: rotation symmetry, isospin symmetry, and so
on. The second reason was that there appeared to be no need for
such theories.

We observe that most of the familiar symmetries are absent at the
Planck scale. There are clearly no conservation laws such as
isospin, and, in the absence of true locality one cannot rotate
any system with respect to a reference system, since they do not
decouple. Constructing counter examples to hidden variable
theories then becomes a lot harder. Also we claim that some
relaxed version of quantum mechanics could be extremely helpful in
bypassing the holographic principle; locality could be restored,
and once again causality could be reconciled with (a weaker
version of) local Lorentz invariance, or general coordinate
invariance. We suspect that such steps may be needed in
constructing logically coherent theories for Planck scale physics.

\ni Thus, one may have one or several motivations for a
reconsideration of ``hidden variable" theories:

\item{$i.$} Einstein's wish for ``reality". At the time this is written, the
quantum mechanical doctrine, according to which all physical
states form a Hilbert space and are controlled by non-commuting
operators, is fully taken for granted in theoretical physics. No
return to a more deterministic description of ``reality" is
considered necessary; to the contrary, string theorists often give
air to their suspicion that the real world is even crazier than
quantum mechanics. One might however complain that the description
of what really constitutes concepts such as space, time, matter,
causality, and the like, is becoming increasingly and
uncomfortably obscure. By some physicists this may be regarded as
an inescapable course of events, with which we shall have to learn
to live, but others such as this author strongly prefer a more
complete description of the notion of reality. We admit however
that at scales relevant to atomic physics, or even the Standard
Model, there is no direct logical inconsistency to be found in
Quantum Mechanics.

\item{$ii.$} ``Quantum Cosmology". How would one describe the `wave function of
the universe'? An extremely important example of a quantum
cosmological model, is a model of gravitating particles in 1 time,
2 space dimensions\ref5. Here, a complete formalism for the
quantum version at first sight seems to be straightforward\ref6,
but when it comes to specifying exact details, one discovers that
we cannot rigorously define what quantum mechanical amplitudes
are, what it means when it is claimed that ``the universe will
collapse with such-and-such probability", what and where the
observers are, what they are made of, and so on. Eventually, one
would have to admit that at cosmological scales, any experiment is
done only once, yielding answers either `yes' or `no', but never
probability distributions. The cosmological wave function is a
dubious notion. Quite conceivably, quantum mechanics as we know it
only refers to repeatable experiments at small time and distance
scales. The true laws of physics are about certainties, not
probabilities.

\ni Note that, since the entire hamiltonian of the universe is
exactly conserved, the ``wave function of the universe" would be
in an exact eigenstate of the hamiltonian, and therefore, the
usual Schr\"odinger equation is less appropriate than the
description of the evolution in the so-called Heisenberg
representation. Quantum states are space-time independent, but
operators may depend on space-time points --- although only if the
location of these space-time points can be defined in a
coordinate-free manner!\fn{Note that, besides energy, also total
momentum and angular momentum of the universe must be conserved
(and they too must be zero).}

\item{iii.} Even at a local scale (i.e. not cosmological), there are problems
that we could attribute to a clash with Quantum Mechanics. Apart
from the question of the cosmological principle, these are:
\itemitem{-} the non-renormalizability of gravity;
\itemitem{-} the fact that the gravitational action (the Einstein-Hilbert
action) is not properly bounded in Euclidean space, while the
Maxwell and Yang-Mills actions are. This is related to the
fundamental instability of the gravitational force.
\itemitem{-}topologically non-trivial quantum fluctuations. They could destroy
the causal coherence of any theory. Perhaps most such fluctuations
may have to be outlawed, as they would also require the boundary
conditions to fluctuate into topologically non-trivial ones.
\itemitem{-} black holes cause the most compelling conflicts with local quantum
mechanics.
\itemitem{-} there still is the mystery of the cosmological constant. It
appears to require a reconsideration not only of physical
principles at the Planck scale, but also at cosmological scales,
since we are dealing here with an infrared divergence that appears
to be cancelled out in a way that requires new physics.

\newsect{2. States and probabilities}
Imagine a universe in which different ``states" are possible. A
state may be characterized any possible way: $$|\j\ket\ =\
\Bigg|\,\vec x_1,\ \vec\s_1,\ \epsffile{man.ps},
\matrix{\hbox{your name and}\cr \hbox{telephone}\cr
\hbox{number}}\Bigg\ket\ ,\eqno(2.1)$$ There are two ways in which
we can introduce the equations of motion, and what we mean by
`state' depends on that. At first sight it seems most natural to
postulate that a state evolves in time. So, at a given time $t$ we
have a state, and there is some equation that tells us what this
state looks like at later (or maybe also at earlier) times $t'$.
We refer to this as the Schr\"odinger picture. But when we realize
that the notion of time may depend on the clocks used, or more
generally, that it requires the introduction of Cauchy surfaces,
we might opt for a different notion of ``state". A state is then
defined as time-independent. The universe is in a particular
state, and in this state all observables may depend on time; the
time variable is then linked to the observable, $\OO(t)$. This is
the Heisenberg picture. It is familiar from quantum mechanics,
where now observables are said to be time-dependent, obeying an
evolution equation of the form $\dd\OO(t)/\dd t = -i[\OO(t),\,H]$.
We will frequently switch between Schr\"odinger and Heisenberg
pictures.

Let us begin with the Schr\"odinger picture. For simplicity then
we take time to be discrete (in the Heisenberg picture, the
question whether time is discrete or continuous would not be so
important). A simple example is a universe that can only be in
three different states, and we have a prescription for the
evolution as follows:\br \cl{\epsffile{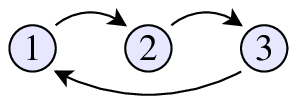}}

This of course is a completely deterministic universe.
Nevertheless, it may be useful to introduce the Hilbert space
spanned by these three states, so as to enable us to handle the
evolution statistically. In this space, the one-time-step
evolution operator would be $$U=\pmatrix{0&0&1 \cr 1&0&0 \cr
0&1&0\cr}\,,\eqno(2.2)$$ a unitary matrix. Suppose we also
consider states of the form
$$|\j\ket=\a|1\ket+\b|2\ket+c|3\ket\,,\eqno(2.3)$$ then after one
time step, we would have $$|\j\ket_{t+1}=\pmatrix{0&0&1 \cr 1&0&0
\cr 0&1&0\cr}|\j\ket_t=U(t,t+1)|\j\ket_t\,.\eqno(2.4)$$ We may
{\it define\/} the probability $P(i)$ for being in a state
$|i\ket$ as $$P(1)=|\a|^2\ ;\quad P(2)=|\b|^2\ ;\quad P(3)=|\g|^2\
,\eqno(2.5)$$ and observe that probability conservation
corresponds to unitarity of the evolution matrix $U$.

Let us now turn to a basis in which $U$ is diagonal:
$$U\ra\pmatrix{1&0&0\cr 0&e^{ 2\pi/3}&0\cr 0&0&e^{-2\pi/3}\cr}\
.\eqno(2.6)$$ One then may write $$U=\exp\big(-iH\d t\big)\
,\eqno(2.7)$$ where the unit time interval $\d t$ will often be
taken to be one. So, we can take as our `hamiltonian':
$$H=\pmatrix{0& &\cr &- 2\pi/3&\cr & &+2\pi/3}\ .\eqno(2.8)$$ Note
that this is the hamiltonian of an atom with a magnetic moment in
a homogeneous magnetic field, although it should of course also be
noted that all eigenvalues of $H$ are only well-defined {\it
modulo\/} integral multiples of $2\pi$.

All that was needed for the above manipulations is that the
deterministic law used as a starting point was time-reversible,
otherwise the matrix $U$ would not have been unitary. Thus, we
make our first important observation:\ref7 {\it any deterministic,
time-reversible system can be described using a quantum mechanical
Hilbert space, where states obey a Schr\"odinger equation, and
where the absolute squares of the coefficients of the wave
functions represent probabilities}. Prototype examples are clocks
that count periodically over $N$ different states --- they can be
mapped onto atoms with spin $j$ in a magnetic field, $N=2j+1$.

The converse is not true: most simple quantum mechanical systems
do not allow a deterministic interpretation. If there would only
be a finite and small number of states, it is rather easy to read
off when a deterministic interpretation may be allowed, but in an
infinite volume limit this is not at all very straightforward, in
particular when non-local transformations are allowed, {\it as
seems to be the case in gravitational physics such as strings and
black holes!} Can a mapping condition be formulated?

If, in a quantum theory, in its Heisenberg formulation, a complete
set of operators $\OO(t)$ can be found that mutually commute at
all times, $$[\OO(t),\,\OO(t')]=0\ ,\quad\forall\,(\,t,\,t')\
,\eqno(2.9)$$ then the theory may be said to be deterministic. A
set of operators is complete if any given set of eigenvalues
$O(t)$ unambiguously specifies a basis element of Hilbert space.
We may take such operators to define expectation values of `truly
existing' observables. In honor of J. Bell, we call these
operators `beables'. The basis generated this way will be called a
{\it primordial\/} basis. It could be that different complete sets
of beables can be found, one set not commuting with another, so
that we have different choices for the primordial basis. In that
case we will have several competing `theories' for the ontological
facts described by our equations.

Operators $P$ that do not commute with the beables of the theory,
such as the evolution operator $U(t,t')$ of Eq.~(2.2), will be
called `changeables'.

Quantum physicists in search of a deterministic theory have the
assignment: find a complete set of beables for the Standard Model,
or a modified version of it. Equivalently, find a primordial
basis. In the model of an atom spinning in a magnetic field, we
succeeded in doing just that. The beables are the matrices
diagonal in the primordial basis
$\big\{|1\ket,|2\ket,|3\ket\big\}$ of Eq.~(2.3).
\newsect{3. `Neutrinos'}

There is a more interesting system, actually realized to some
extent in the real world, for which a  primordial basis can be
constructed.\ref4 Consider massless, non-interacting chiral
fermions in four space-time dimensions. We can think of neutrinos,
although of course real neutrinos deviate slightly from the ideal
model described here.

First, take the first-quantized theory. The hamiltonian for a
Dirac particle is $$H={\vec \a}\cdot{\vec p}+\b
m\,,\qquad\{\a_i,\,\a_j\}=2\d_{ij}\,,\quad \{\a_i,\,\b\}=0\,,\quad
\b^2=1\,.\eqno(3.1)$$ Taking $m=0$, we can limit ourselves to the
subspace projected out by the operator $\half(1+\g_5)$, at which
point the Dirac matrices become two-dimensional. The Dirac
equation then reads $$H={\vec \s}\cdot{\vec p}\,,\eqno(3.2)$$
where $\s_{1,\,2,\,3}$ are the Pauli matrices. We now consider the
following candidates for `beables': $$\big\{\,\hat p\ ,\quad \hat
p\cdot {\vec \s}\ ,\quad \hat p\cdot\vec  x(t)\,\big\}\
,\eqno(3.3)$$ where $\hat p$ stands for $\pm\vec p/|p|$, with the
sign such that $\hat p_x>0$. We do {\it not\/} directly specify
the sign of $\vec p$.

Writing $p_j=-i\part{ }{x_j}$, one readily checks that these three
operators commute, and that they continue to do so at all times.
Indeed, the first two are constants of the motion, whereas the
last one evolves into $$\hat p\cdot\vec x(t)=\hat p\cdot\vec
x(0)+\hat p\cdot\vec\s\,t\,.\eqno(3.4)$$

The fact that these operators form a complete set is also easy to
verify: in momentum space, $\hat p$ determines the orientation;
let us take this to be the $z$ direction. Then, in momentum space,
the absolute value of $p$, as well as its sign, are identified
with its $z$-component, and it is governed by the operator
$i\pa/\pa p_z=x_z=\hat p\cdot\vec x$. The spin is defined in the
$z$-direction by $\hat p\cdot\vec\s$.

Mathematically, these equations appear to describe a {\it plane},
or a flat membrane, moving in orthogonal direction with the speed
of light. Given the orientation (without its sign) $\hat p$, the
coordinate $\hat p\cdot\vec x$ describes its distance from the
origin, and the variable $\hat p\cdot\vec\s$ specifies in which of
the two possible orthogonal directions the membrane is moving.
Note that, indeed, this operator flips sign under $180^\circ$
rotations, as it is required for a spin $\half$ representation.
This, one could argue, is what a neutrino really is: a flat
membrane moving in the orthogonal direction with the speed of
light. But we'll return to that later: the theory can be further
improved (see the end of Sect.~4).

We do note, of course, that in the description of a single
neutrino, the hamiltonian is not bounded from below, as one would
require. In this very special model, there is a remedy to this,
and it is precisely Dirac's second quantization procedure. We
consider a space with an infinite number of these membranes,
running in all of the infinitely many possible directions $\hat
p\cdot\vec\s$. In order to get the situation under control, we
introduce a temporary cut-off: in each of the infinitely many
possible directions $\hat p$, we assume that the membranes sit in
a discrete lattice of possible positions. The lattice length $a$
may be as small as we please. Furthermore, consider a box with
length $L$, being as large as we please. The first-quantized
neutrino then has a finite number of energy levels, between
$-\pi/a$ and $+\pi/a$. The state we call `vacuum state', has all
negative energy levels filled and all positive energy levels
empty. All excited states now have positive energy. Since the
Dirac particles do not interact, their numbers are exactly
conserved, and the collection of all observables (3.3) for all
Dirac particles still correspond to mutually commuting operators.

In this very special model we thus succeed in producing a complete
primordial basis, generated by operators that commute with one
another at all times (beables), whereas the hamiltonian is bounded
from below. We consider this to be an existence proof, but it
would be more satisfying if we could have produced a less trivial
model. Unfortunately, our representation of neutrinos as infinite,
strictly flat membranes, appears to be impossible to generalize so
as to introduce mass terms and/or interactions.

Further attempts at obtaining more realistic models which are
fundamentally quantum mechanical yet allow for a deterministic
interpretation failed because it did not appear to be possible to
create a hamiltonian that is bounded below, so that a very special
state can be selected, being the lowest eigenstate, which can be
identified unambiguously as the vacuum state. In the Schr\"odinger
picture, two classes of models may be considered: the ones with
{\it continuous time}, and the ones with {\it discrete time}. If
time is continuous, and a set of beables $q_i(t)$ is found, a
natural choice for the hamiltonian would be $$H=\sum_i p_i\cdot
f_i({\bf q})\ ;\quad \dot q_i=f_i({\bf q})\ ,\eqno(3.5)$$ where
$p_i$ are the ordinary momentum operators associated to $q_i$. We
see immediately that, in spite of the quantum mechanical notation,
the $q_i$ evolve in a deterministic manner. But we also see that,
regardless the choice for the functions $f_i$, this hamiltonian
can never be bounded below.

Time does not have to be continuous. It would suffice if a set of
beables could be defined to form a sufficiently dense lattice in
space-time, and this brings us to cellular automaton models. In
these models, only a finite number of possible states is needed in
a given spacelike volume. One might hope that then also the
Hamilton density would be a finite-dimensional matrix, so that the
existence of a ground state might follow, in certain cases. Again,
however, there is a problem. Let now $$q_i(t+1)=f_i\big({\bf
q}(t)\big)\,,\eqno(3.6)$$ then this defines uniquely the evolution
operator over integer time steps: $$U(t+1,t)=\ex{-i
H}\,,\eqno(3.7)$$ but then this defines the eigenvalues of $H$
only {\it modulo\/} $2\pi$. Again, the notion of a lowest
eigenstate is questionable.

In the Heisenberg picture, the dimensionality of a limit cycle
does not change if we replace the time variable by one with
smaller time steps, or even a continuous time. Working with a
continuous time variable then has the advantage that the
associated operator, the hamiltonian, is unambiguous in that case.

The neutrino example of this section does show that in some cases
a bounded hamiltonian may yet exist (here, it is the continuum
limit that singles out a special choice for $H$ with unambiguous
eigenvalues), so one may hope that more is possible, but something
drastically new may be needed.

\newsect{4. Information loss}

The new ingredient needed might be information loss\ref8. At first
sight this is surprising. One would have thought that, with
information loss, the evolution operator will no longer be
unitary, and hence no quantum mechanical interpretation is
allowed. Consider however a model universe with 4 elements. They
will be indicated not by Dirac brackets, but as (1), (2), (3) and
(4). The evolution law is as depicted in Fig.~1.

\midinsert\cl{\epsffile{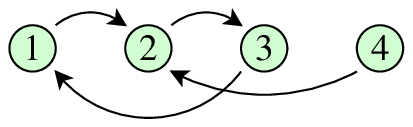}}  \cl{Fig.~1. Transition rule
with information loss.}  \endinsert

If we would associate basis elements of a Hilbert space to each of
these states, the evolution operator would come out as
$$U(t+1,t)=\pmatrix{0&0&1&0\cr 1&0&0&1\cr 0&1&0&0\cr 0&0&0&0}\
,\eqno(4.1)$$ and this would not be a unitary operator. Of course,
the reason why the operator is not unitary is that the evolution
rule (2.1) is not time reversible. After a short lapse of time,
only states (1), (2) and (3) can be reached. In this simple
example, it is clear that one should simply erase state (4), and
treat the upper left $3\times3$ part of Eq. (4.1) as the unitary
evolution matrix. Thus, the quantum system corresponding to the
new evolution law  is three-dimensional, not four-dimensional, and
it seems to be trivial to restore time-reversibility and
unitarity.

In more complicated non-time-reversible evolving systems, however,
the `genuine' quantum states and the false ones (the ones that
cannot be reached from the far past) are actually quite difficult
to distinguish. For this reason, we introduce the notion of {\it
equivalence classes}. Two states are called equivalent if, after
some finite time interval, they evolve into the same state. The
system described above has three equivalence classes,
$$E_1=\{(1),\, (4)\}\ ,\quad E_2=\{(2)\}\ ,\quad
E_3=\{(3)\}\,.\eqno(4.2)$$ {\it Quantum states will now be
identified with these equivalence classes.} In our example, in the
Schr\"odinger picture, we have three basis elements, $|1\ket=E_1;\
|2\ket=E_2;\ |3\ket=E_3$. {\it In terms of these objects} $E_1,\
E_2,\ E_3$, one has an evolution operator  $$U=\pmatrix{0&0&1\cr
1&0&0\cr 0&1&0}\,,\eqno(4.3)$$  and a hamiltonian operator $H$ can
be defined such that $U=e^{-iH}$. Our model universe would be in
an eigenstate of this hamiltonian.

An extreme example of a situation where equivalence classes must
occur is the black hole. We imagine a theory with classical (that
is, not quantum mechanical) general relativistic field equations.
Classical black holes may result as solutions. Since this
classical system surely has information loss, the equivalence
classes $E_i$ will each comprise all possible initial states that
result in the same black hole (with the same mass, charge and
angular momentum) after collapse. It is important to realize that
the equivalence classes may be much smaller than the classes of
primordial `states'. Their definition is not local, in the sense
that two states that differ at different locations may belong to
the same equivalence class. One might suspect that this could
explain the apparent need for non-locality in conventional
attempts at constructing hidden variable theories.

The physical distinction between theories with information loss
and theories without information loss is not very clear in models
with a small number of distinct states such as the example given
above. After all, one may simply ignore the `unreachable' states
and notice that the universe ends up in a limit cycle that is
indistinguishable from what a universe without information loss
would do. It is the fact that the true universe is far too large
ever to end up in a limit cycle, and the fact that we wish to
understand small regions of this universe, nearly but not quite
decoupled from the rest of the world, which make the introduction
of information loss and equivalence classes non-trivial. The most
acute problem to be addressed is how to create a hamiltonian that
can be viewed as the integral over a hamiltonian density that is
bounded below and has (more or less) local commutation rules.

What we address next is the question how to construct the laws of
physics for a universe that is essentially open, {\it i.e.} it
consists of smaller parts glued together. These smaller parts form
an infinite 3 (space-)dimensional world. We begin with gluing two
small pieces together.

If left alone, a tiny segment of the universe may be assumed to
enter into a limit cycle. In fact, it may have several different
cycles to choose from, depending on the initial state. In the
Schr\"odinger picture, each of these cycles forms a sequence of a
large number of states, and there are numerous different energy
eigenstates that the system can choose from. But, as stated
before, if time is not an extrinsically defined coordinate, it is
meaningless to consider time dependence. We then prefer the
Heisenberg picture, where we have only one state for each cycle.
In this state it is the observables that take a periodic sequence
of values, but only if time could be defined at all. In any case,
the dimensionality of Hilbert space is then determined by the
number of different possible cycles.

If we have two adjacent segments of the universe, the {\it
relative} time coordinate is well-defined and important. It may or
may not be defined as an arbitrary real number. A way to introduce
coupling is as follows. We introduce a cyclic time coordinate on
each segment (which could be seen as a reintroduction of the
Schr\"odinger picture). Now assume that the relative speed of the
time evolution of the two adjacent segments is determined by local
dynamics. Using general relativistic notation, one would have
$${\dd\over\dd
t}\big(|\j_1\ket|\j_2\ket\big)=\Big(\sqrt{g^{00}(1)}\,{\pa\over\pa
t}|\j_1\ket\Big)|\j_2\ket+\sqrt{g^{00}(2)}\ |\j_1\ket{\pa\over\pa
t}|\j_2\ket\,.\eqno(4.4)$$ Here, $\sqrt{g^{00}}$ may be
interpreted as the `gravitational potential field', and the ratio
$\sqrt{g^{00}(2)}/\sqrt{g^{00}(1)}$ is the `gravitational field
strength'.

 Take the case that each segment has just one limit
cycle. Each now has a periodic `time' variable $q\in [\,0,\,1\,)$,
but we also take an external time variable $t$, so we have
$q_1(t)$ and $q_2(t)$.  Let the undisturbed time derivatives be
$$\dd q_i/\dd t \equiv \dot q_i(t)=v_i\,,\eqno(4.5)$$ so that the
(undisturbed) periods are $T_i=1/v_i$. The really relevant
quantity is the ratio $$\dd q_1(t)/\dd t : \dd q_2(t)/\dd t = \dd
q_1/\dd q_2\ .$$  In Fig.~2, this is the slope of the trajectories
for the solutions. \midinsert\cl{\epsffile{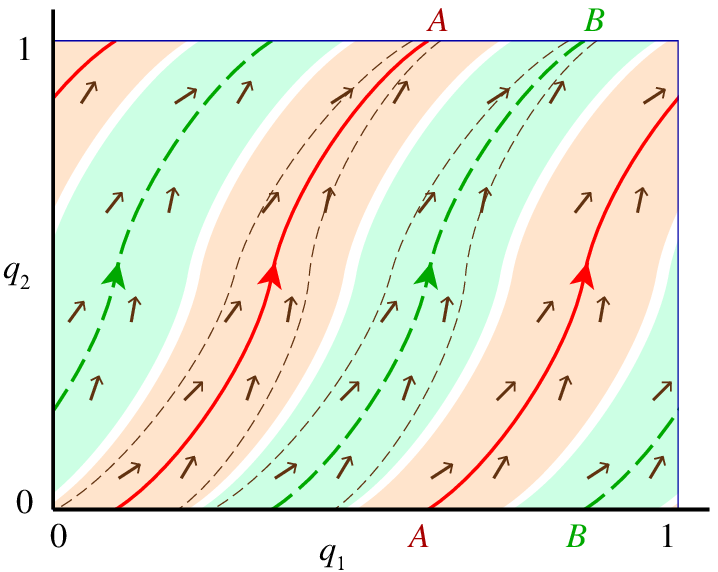}} \Narrower
Fig.~2. Flow chart of a continuum model with two periodic
variables, $q_1$ and $q_2$. In this example, there are two stable
limit cycles, $A$ and $B$, representing the two `quantum states'
of this `universe'. In between, there are two orbits that would be
stable in the time-reversed model. \endinsert \ni

 The Schr\"odinger Hilbert space is
spanned by the states $|q_1,\,q_2\ket$, and our formal hamiltonian
is $$H=v_1p_1+v_2p_2\,;\qquad p_j=-i\pa/\pa q_j\,.\eqno(4.6)$$ In
this case, even the zero-energy states span an infinite Hilbert
space, so, in the Heisenberg picture, the product universe has an
infinity of possible states.

Information loss is now introduced by adding a tiny perturbation
that turns the flow equations into a non-Jacobian one: $$v_1\ra
v_1^0+\e f(q_1,q_2)\,;\qquad v_2\ra v_2^0+\e
g(q_1,q_2)\,.\eqno(4.7)$$ The effect of these extra terms can vary
a lot, but in the generic case, one expects the following
(assuming $\e$ to be a sufficiently tiny number):

Let the ratio $v_1^0/v_2^0$ be sufficiently close to a rational
number $N_1/N_2$. Then, at specially chosen initial conditions
there may be periodic orbits, with period
$$P=v_1^0/N_1=v_2^0/N_2\,,\eqno(4.8)$$ where now $v_1^0$ and
$v_2^0$ have been tuned to exactly match the rational ratio ---
possible deviations are absorbed into the perturbation terms.
Nearby these stable orbits, there are non-periodic orbits, which
in general will converge into any one of the stable ones, see
Fig.~2. After a sufficiently large lapse of time, we will always
be in one of the stable orbits, and all information concerning the
extent to which the initial state did depart from the stable
orbit, is washed out. Of course, this only happens if the Jacobian
of the evolution, the quantity $\sum_i(\pa/\pa q_i)\dot q_i$,
departs from unity. Information loss of this sort normally does
not occur in ordinary particle physics, although of course it is
commonplace in macroscopic physics, such as the flow of liquids
with viscosity (see Sect.~6).

The stable orbits now represent our new equivalence classes (note
that, under time reversal, there are other stable orbits in
between the previous ones). Most importantly, we find that the
equivalence classes will form a discrete set, in a model of this
sort, most often just a finite set, so that, back in the
Heisenberg picture, our `universe' will be just in a finite number
of distinct quantum states.

Generalizing this model to the case of more than two periodic
degrees of freedom is straightforward. We see that, if the flow
equations are allowed to be sufficiently generic (no constraints
anywhere on the values of the Jacobians), then distinct stable
limit orbits will arise. There is only one parameter that remains
continuous, which is the global time coordinate. If we insert
$H|\j\ket=0$ for the entire universe, then the global time
coordinate is no longer physically meaningful, as it obtains the
status of an unobservable gauge degree of freedom.

In the above models, what we call `quantum states', coincides with
Poincar\'e limit cycles of the universe. We repeat, just because
our model universes are so small, we were able to identify these.
When we glue tiny universes together to obtain larger and hence
more interesting models, we get much longer Poincar\'e cycles, but
also much more of them. Eventually, in practice, sooner or later,
one has to abandon the hope of describing complete Poincar\'e
cycles, and replace them by the more practical definitions of
equivalence classes. At that point, when one combines mututally
weakly interacting universes, the effective quantum states are
just multiplied into product Hilbert spaces.

Our introduction of information loss and equivalence classes sheds
new light on the neutrino model introduced in Sect.~3. In that
section, we concluded that neutrinos seem to `be' infinite, flat,
sheets, which may eventually become untenable when space-time
curvature and other interactions or mass terms are taken into
account. Now, we have another option. The sheets are not the
primordial beables but they are the equivalence classes. We could
have that neutrinos are more-or-less conventional, but classical,
point particles, with auxiliary velocity vectors $\bf p$ of unit
length. The rules of motion are now such that the velocity in the
direction of $\bf p$ is rigorously fixed to be $c$, but the
velocities in the transverse directions are chaotic and not
re-traceable due to information loss. In such a picture it is no
longer impossible to imagine tiny deviations from the rule to
incorporate interactions and tiny mass terms.

\newsect{5. Harmonic and anharmonic oscillators.}

What we have so far is a strategy. We still have the question how
a model, either with or without information loss, can emerge in
such a way that the total hamiltonian is the integral of a
Hamilton density bounded from below, so that we can understand the
chaotic nature of our vacuum. The neutrino model was one very
special case. Now let us concentrate on the most elementary
building block for bosonic fields in Nature: the harmonic
oscillator, with possible disturbances.

A {\it classical\/} harmonic oscillator may be described by the
equations $$\dot x= y\ ;\quad \dot y=-x\ .\eqno(5.1)$$ The
(Schr\"odinger) states are then all sets $|x,\,y\ket$, and the
classical equation (5.1) is generated by the `hamiltonian'
$$H=y\,p_x-x\,p_y\,.\eqno(5.2)$$ This, of course, is not bounded
below, the price to be paid for writing a classical system quantum
mechanically.

We may rewrite however,
$$\eqalign{H&=\quart(y+p_x)^2-\quart(y-p_x)^2-\quart(x+p_y)^2+\quart(x-p_y)^2\cr
&=H_1-H_2\cr H_1&=\half P_1^2+\half Q_1^2\ ;\qquad H_2=\half
P_2^2+\half Q_2^2\ ;}\eqno(5.3)$$ with $$\eqalign{
P_1&=\ffract1{\sqrt2}\,(p_x+y)\ ;\qquad
P_2=\ffract1{\sqrt2}\,(x+p_y)\ ;\cr
Q_1&=\ffract1{\sqrt2}\,(x-p_y)\ ;\qquad
Q_2=\ffract1{\sqrt2}\,(y-p_x)\ .}\eqno(5.4)$$ The new variables
$P_i$ and $Q_i$ obey the usual commutation rules:
$$[P_i,\,Q_j]=-i\d_{ij}\ ;\qquad [P_i,\,P_j]=[Q_i,\,Q_j]=0\
,\eqno(5.5)$$ so that the two parts of the hamiltonian (5.3)
commute: $[H_1,\,H_2]=0$. Thus, we found that although the
classical harmonic oscillator has an unbounded quantum
hamiltonian, we can decompose it into {\it two\/} genuine quantum
hamiltonians, both of the familiar harmonic oscillator type and
both bounded from below, but with a minus sign in between. All we
now have to do is to `postulate' that, say,
$H_2|\j\ket=\half\hbar\w$ is imposed as a constraint equation on
all states. Then we have a quantum harmonic oscillator.

This, however, is not the true solution to our problem. The
splitting (5.3) is far too arbitrary. Where does the constraint
$H_2=\half$ come from, and how can it survive interactions? How do
we couple two or more such oscillators without spoiling the
constraint? Let us analyze the reason why the splitting (5.3) is
possible.

The classical harmonic oscillator has two conserved operators,
besides the hamiltonian itself. These are:\br --- the radius $r$
of the orbit: $\r^2=x^2+y^2$, and\br --- the {\it dilatation
operator} $D=x\,p_x+y\,p_y-i$; classically, after all, the
periodic solutions of the oscillator behave identically after a
scale transformation $x,\,y\ra \l x,\,\l y\,;\ p_x\,,p_y\ra
\l^{-1}p_x,\,\l^{-1}p_y$. We easily check that $$[H,\,x^2+y^2]=0\
;\qquad [H,\,D]=0\ .\eqno(5.6)$$ It is due to these operators that
we can construct $H_1$ and $H_2$:
$$\eqalign{H_1&={1\over4\r^2}(\r^2+H)^2+{1\over4\r^2}(D+i)^2\ ;\cr
H_2&={1\over4\r^2}(\r^2-H)^2+{1\over4\r^2}(D+i)^2\ .}\eqno(5.7)$$
We see immediately that $H_1$ and $H_2$ commute, since $D$ and
$\r$ commute with $H$ (they do not commute with each other, but
that does no harm), and we see that $H=H_1-H_2$.

But we also see that, in this notation, the construction is fairly
arbitrary. The contribution of the dilatation operator is
unnecessary. We may perform the more interesting splitting:
$$\eqalign{\hbox{if}\qquad[\r,\,H]=0\qquad& \hbox{then}\qquad
H=H_1-H_2\cr\qquad& H_1={1\over4\r^2}(\r^2+H)^2\ ;\cr
 & H_2={1\over4\r^2}(\r^2-H)^2\ .}\eqno(5.8)$$

Now this generalizes to any function $\r^2$ of the coordinates $x$
and $y$ that happens to be conserved, $[\r^2,\,H]=0$. So, in
general, let $$H={\bf p \cdot\,f(q)}\ ;\qquad [H,\,\r^2({\bf
q})]=0\ ,\eqno(5.9)$$ and $$\eqalign{H=H_1-H_2\ &;\cr
H_{1,2}={1\over4\r^2}(\r^2\pm H)^2\ &;\qquad [H_1,\,H_2]=0\ .}
\eqno(5.10)$$ Then introduce as a constraint: $$H_2|\j\ket\ra 0\ ,
\eqno(5.11)$$ this implies that $$H\ra H_1\ra\r^2\ge
0\,.\eqno(5.12)$$

 \midinsert\cl{\epsffile{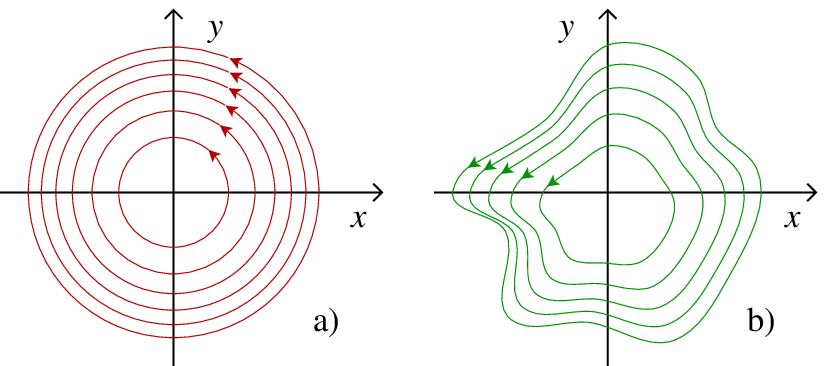}} \Narrower Fig.~3.
Stable orbits, a) for the harmonic oscillator, b) an anhormonic
oscillator. After switching on a dissipative term, the regions in
between these trajectories will have only non-periodic solutions,
tending towards the stable attractors. \Endnarrower\endinsert

The subclass of states obeying the constraint (5.11) obeys the
Schr\"odinger equation ${\dd\over\dd t}|\j\ket=-iH_1|\j\ket$, with
a hamiltonian $H_1$ bounded from below. {\it But now we can
motivate the constraint by introducing information loss!} This
goes as follows. Let us assume that, in a deterministic system,
there are stable orbits, separated by regions where solutions are
non-periodic, attracted to the stable attractors. First, we mimic
the system in terms of a non-dissipative model, where all orbits
would be stable. It is then described by Eq.~(5.9). Let the
periods of these orbits be functions $T(\r)$ of $\r$. The
constraint (5.11), (5.12) implies $H=\r^2$, or $$\ex{-iHT}|\j\ket
= |\j\ket\ .\eqno(5.13)$$ These are the states for which the
periods $T(\r)$ obey $$\r^2T(\r)=2\pi n\ ,\qquad n\in  \Bbb Z\ .
\eqno(5.14)$$ Thus, the constraint appears to correspond  to
limiting oneself to the stable orbits only. Note that, with
Eq.~(5.14), the hamiltonian can obtain any kind of eigenvalue
spectrum, as opposed to the equidistant lines of the harmonic
oscillator.

\newsect{6. Conclusions}
In this lecture we investigated {\it classical, deterministic,
dissipative\/} models, and we found that, in general, they develop
distinct stable orbits. The mathematics for analyzing these models
requires that we first introduce non-dissipative equations, which
allow a formalism using quantum mechanical notation, but, without
dissipation, it cannot be understood why the hamiltonian would be
bounded from below. Then we find that dissipation imposes
constraints on the solutions, which appear to provide bounded
hamiltonians. It is remarkable that dissipation also leads to an
apparent {\it quantization\/} of the orbits, and this quantization
indeed resembles the quantum structure seen in the real world.

The next step, yet to be taken, is to couple infinite numbers of
dissipating oscillators to form models of quantum field theories.
This may appear to be a very difficult task, but we do notice that
in classical general relativity black hole formation is
inevitable, and black holes indeed absorb information. This would
imply that the distance scale at which dissipation plays a role
must be the Planck scale.

At scales between the Standard Model and the Planck scale, the
introduction of dissipation would be a new approach. Perhaps we
can find models resembling Navier-Stokes liquids with viscosity.
In non-relativistic models, the dimensionality of a viscosity
$\eta$ is given by $$\big[\eta/\r]=[{\rm cm}^2/{\rm
sec}]\,,\eqno(6.1)$$ where $\r$ is the mass density of the fluid.
In a relativistic theory, where there is a fixed unit of velocity
$c=1$, the cm and the sec have the same dimensionality, so now
$\eta/\r$ has the dimensionality of a length. It is tempting to
take this to be the Planck length. We may take strictly continuous
fields, which however at distance scales $\ell$ small compared to
the Planck scale are totally controlled by viscosity. There, at
small Reynolds number\ref9,  $$ R=\r u\ell/\eta\,,\eqno(6.2)$$
where $u$ are the typical velocities, the field distributions show
no further structure, but at distance scales large compared to the
Planck scale, one may expect `turbulence', {\it i.e.}, chaotic
behaviour, for which we propose the introduction of apparently
quantum mechanical techniques in order to describe the statistics.
One then may invoke the renormalization group in order to reach
the length scales of the Standard Model, and make contact with the
real world.

It is far too early to ask for tangible results and firm testable
predictions of the approach that we have in mind. A very indirect
prediction may perhaps be made. We conjecture that the apparently
quantum mechanical nature of our world is due to the statistics of
fluctuations that occur at the Planck scale, in terms of a regime
of completely deterministic dynamics. This would entail that all
quantum mechanical effects should be reproducible in some
deterministic model, including all machinations with what is
usually called a `quantum computer'. As is well-known, quantum
computers, if they can be constructed, will be able to do
computations no ordinary computer can accomplish. This would be a
contradiction with our claim that it can be mimicked using
ordinary computers. However, we are unable to mimic any quantum
system we like. Interactions are essential and unavoidable. It is
also these interactions that cause unwanted decoherence in a
quantum computer. Experimenters are trying to create devices in
which the ideal situation is approached as well as possible. I now
claim that it will be impossible to shield these devices from the
unwanted interactions, so that the ideal quantum computer can
never be built. More precisely: \Narrower {\it No quantum computer
can ever be built that can outperform a classical computer if the
latter would have its components and processing speed scaled to
Planck units.}\Endnarrower

\ni Because the Planck units are extremely tiny, this still leaves
lots of room for quantum computers to do miracles, but eventually,
there will be a limit. An ideal quantum computer that would
consist of millions of parts, could, in principle, factor integers
with millions of digits into prime numbers. It appears that
present programs can factor a number of $N$ digits using memories
of the order of $10^{\sqrt{N\log N}}$ cells, in $10^{\sqrt{N\log
N}}$ steps. Perhaps a reasonable computer takes $10^{120}$ Planck
volumes; that would limit the factorizable numbers to $10^{4000}$
or so. Thus we predict that even a quantum computer will not be
able to exceed such limits in practice.

\newsect{References}

\item{1.} G.~'t~Hooft, {\it The Holographic Principle}, Opening
Lecture, Erice, August 1999. See also:\br
 G.~'t~Hooft, {\it Dimensional reduction in quantum gravity.} In {\it
Salamfestschrift: a collection of talks}, World Scientific Series
in 20th Century Physics, vol.~{\bf 4}, Eds.~A.~Ali, J.~Ellis and
S.~Randjbar-Daemi (World Scientific, 1993), THU-93/26,
gr-qc/9310026;  {\it Black holes and the dimensionality of
space-time}, in Proceedings of the Symposium ``The Oskar Klein
Centenary'', 19-21 Sept. 1994, Stockholm, Sweden. Ed. U.
Lindstr\"om, World Scientific 1995,  p.~122; L.~Susskind,  {\it
J.~Math.~Phys. \bf 36} (1995) 6377, hep-th/9409089.
\item{2.}M.B.~Green, J.H.~Schwarz and E.~Witten, {\it Superstring Theory},  Cambridge
     Univ. Press.
\item{3.}E.~Witten, {\it Anti de Sitter Space and holography}, hep-th/9802150;
 J.~Maldacena, {\it The large $N$ Limit of superconformal field theories and supergravity}, hep-th/9711020;
 T.~Banks et al, {\it Schwarzschild Black Holes from Matrix Theory}, hep-th/9709091;
 K.~Sken\-deris, {\it Black holes and branes in string theory}, SPIN-1998/17, hep-th/9901050.
\item{4.}G.~'t~Hooft,{\it Quantum Gravity as a Dissipative Deterministic System}, SPIN-1999/07, gr-qc/9903084;
 Class.~Quant.~Grav. 16 (1999) 3263.
\item{5.}A.~Staruszkiewicz, {\it Acta Phys. Polon. \bf 24} (1963) 734;
     S.~Giddings, J.~Abbott and K.~Kuchar, {\it Gen.~Rel. and Grav. \bf  16}  (1984)
     751; S.~Deser, R.~Jackiw and G.~'t~Hooft, {\it Ann.~Phys. \bf 152} (1984) 220;
J.R.~Gott, and M.~Alpert, {\it Gen.~Rel.~Grav. \bf 16} (1984) 243;
 J.R.~Gott, {\it Phys.~Rev.~Lett.} {\bf 66} (1991) 1126;
 S.~Deser, R.~Jackiw and G.~'t~Hooft, {\it Phys.~Rev.~Lett.} {\bf 68} (1992) 267;
S.M.~Carroll, E.~Farhi and A.H.~Guth, {\it Phys.~Rev.~Lett.} {\bf 68} (1992) 263;
 G.~'t~Hooft,  {\it Class.~Quantum Grav.
 \bf 9} (1992) 1335.
\item{6.} A.~Achucarro and P.K.~Townsend, {\it Phys.~Lett. \bf B180} (1986) 89;
E.~Witten, {\it Nucl.~Phys.} {\bf B311} (1988) 46; S.~Carlip, {\it
Nucl.~Phys.} {\bf B324} (1989) 106, and in: "Physics,  Geometry
and Topology", NATO ASI series B, Physics, Vol. {\bf 238}, H.C.~
Lee  ed.,  Plenum 1990, p.~541; {\it Six ways to quantize
(2+1)-dimensional gravity}, Davis Preprint UCD-93-15,
gr-qc/9305020; G.~'t Hooft, {\it Class.~Quantum Grav.} {\bf 10}
(1993) 1023, {\it ibid.} {\bf 10} (1993) S79; {\it {\it
Nucl.~Phys.}} {\bf B30} (Proc.~Suppl.) (1993) 200; {\it
Class.~Quantum Grav.\bf 13} 1023, gr-qc/9601014.
\item{7.} G. 't Hooft, {\it Nucl.~Phys.\bf B342 }(1990) 471.
\item{8.} G.~'t~Hooft, {\it Found.~Phys. letters \bf 10} (1997) 105, quant-ph/9612018.
\item{9.} See e.g. L.D.~Landau and E.M.~Lifshitz, Course of Theoretical Physics,
Vol 6, {\it  Fluid Mechanics}, Pergamon Press, Oxford 1959.

\bye